\documentclass[a4paper,fleqn]{cas-sc}

\usepackage[none]{hyphenat}
\usepackage{graphicx}
\usepackage{float}
\usepackage{amsfonts}
\usepackage{amsmath}
\usepackage{amssymb}
\usepackage{mathrsfs} 
\usepackage{stmaryrd}
\usepackage{comment}
\usepackage{subcaption,siunitx,booktabs}

\usepackage[authoryear]{natbib}
\usepackage{tikz}
\usepackage{circuitikz}
\usepackage{pgfplots}
\pgfplotsset{compat=1.18}
\usetikzlibrary{shapes,shadings,arrows,shapes.multipart,positioning,decorations.text,fit}

\begin{document}
\let\WriteBookmarks\relax
\def\floatpagepagefraction{1}
\def\textpagefraction{.001}

\shorttitle{}    

\shortauthors{}  

\title [mode = title]{Schumann Resonances as a tool to constrain the depth of Titan’s buried water ocean: 
Re-assessment of Huygens observations and preparation of the EFIELD/Dragonfly experiment}

\author[inst1]{\href{mailto:paul.lagouanelle@latmos.ipsl.fr}{Paul Lagouanelle}}
 \author[inst1]{\href{mailto:alice.legall@latmos.ipsl.fr}{Alice Le Gall}}

\affiliation[inst1]{organization={LATMOS/IPSL},
            addressline={Université Paris-Saclay, UVSQ, Sorbonne Université, CNRS}, 
            postcode={78280},
            city={Guyancourt},
            country={France}}

\begin{abstract}
Among the lines of evidence for a buried ocean on Titan is the possible detection, in 2005, by the Permittivity, Wave and Altimetry (PWA) analyzer on board the ESA Huygens probe of Schumann-like Resonances (SR). SR are Extremely Low Frequency electromagnetic waves resonating between two electrically conductive layers. On Titan, it has been proposed that they propagate between the moon’s ionosphere and a salty subsurface water ocean. Their characterization by electric field sensors can provide constraints on Titan’s cavity characteristics and in particular on the depth of Titan’s ocean which is key to better assess Titan’s habitability. For this work we have developed a numerical model of Titan’s electromagnetic cavity as well as a surrogate model to conduct simulations and sensitivity analyses at a low computational cost. This surrogate model is used both to re-assess PWA/Huygens measurements and to predict the future performance of the EFIELD experiment on board the NASA Dragonfly mission. We demonstrate that the PWA/Huygens measurements, in particular due to their low spectral resolution, do not bring any meaningful constraint on Titan’s ocean depth. On the other hand, the finer resolution of the EFIELD experiment and its ability to capture several harmonics of SR should provide more robust constraints on Titan’s internal structure, especially if the electrical properties of the ice crust and the atmosphere can be better constrained.
\end{abstract}


\begin{highlights}
\item Surrogate modeling of Titan's planetary cavity for Schumann resonances
\item Re-assessment of Huygens observations to constrain the thickness of Titan's ice crust
\item Potential performances of the EFIELD experiment on board Dragonfly for estimating the thickness of the ice crust
\end{highlights}

\begin{keywords}
Titan, interior \sep
Instrumentation \sep
Ionospheres \sep
Radio observations \sep
Titan, atmosphere
\end{keywords}

\maketitle

\section{Introduction}
\label{sec:introduction}

Several lines of evidence point to the presence of a global water ice ocean in Titan’s interior. The strongest evidence arises from the investigation of the tidal variations of Titan’s gravity fields inferred from Cassini flybys of the satellite \citep{iess2012tides,durante2019titan}. Indeed, the tidal Love number $k_2$ of $0.62$ derived by \cite{durante2019titan} is compatible with a high-density ocean while a recent re-assessment of $k_2$ ($0.375$) points to a low-density water or ammonia ocean \citep{goossens2024low}. Titan’s measured obliquity of $\sim 0.3^{\circ}$ \citep{stiles2008determining,meriggiola2016rotational} is also significantly larger than the value expected for an entirely solid object and therefore suggests a decoupling between the outer ice shell and the interior of Titan \citep{baland2011titan,baland2014titan,bills2008forced,bills2011rotational}. Based on both the values of \cite{durante2019titan}’s $k_2$ and of the obliquity, \cite{baland2014titan} estimate that the outer ice shell of Titan is at least \SI{40}{\kilo\meter} and at most \SI{170}{\kilo\meter} thick consistent with the results published in \cite{kronrod2020matching} which test a wide range of internal structure models for Titan including thermal considerations. 

Another, possible evidence for an internal ocean on Titan is the detection of ELF (Extremely Low Frequency) waves by the PWA/HASI (Permittivity, Waves and Altimetry analyzer, part of the Huygens Atmospheric Structure Instrument) experiment on board the Huygens interpreted as Schumann Resonances \citep{beghin2012analytic}. Schumann resonances (SR) are a set of ELF electromagnetic propagation modes that can develop in a planetary cavity excited with a broadband electromagnetic source \citep{schumann1952strahlungslosen}. On Earth, these modes are generated by lightning discharges and propagate between the ionosphere and the surface. Theoretically, SR could be observed on other planets and serve as a tool to obtain information on the planetary cavities, in particular on their dimensions \citep{simoes2007phd,simoes2008schumann,simoes2008electromagnetic}.

If SR exist on Titan, there are probably not triggered by lightning as such activity was never observed and is not expected to be common on Titan \citep{lorenz1997lightning}. In addition, the surface of Titan being very poorly conductive \citep{grard2006electric,hamelin2016electrical}, it cannot act as the lower boundary of the resonant cavity which instead must be an internal electrically conductive layer. \cite{beghin2012analytic} propose that SR on Titan could be excited by interactions with Saturn’s magnetosphere and that the signal detected at $\sim\SI{36}{\Hz}$ by PWA/HASI is the second harmonic of a SR propagating between Titan’s ionized atmospheric layer (at $\sim60-\SI{70}{\kilo\meter}$ altitude) and a buried salty ocean lying at a depth encompassed between $40$ and \SI{80}{\km}. However, this interpretation is still debated as the \SI{36}{\Hz} line may actually be due to mechanical oscillations of the booms on which the PWA/HASI electrodes were installed or of other parts of the Huygens probe \citep{lorenz2020schumann}. 

Nevertheless, if SR occur on Titan, their detection and characterization would place new and more robust constraints on the buried ocean. In particular, knowing more accurately its depth is key to estimate the likeliness of exchange between the ocean and the surface and therefore to assess Titan’s habitability and astrobiological potential. That is the reason why the forthcoming mission to Titan, Dragonfly (NASA), will embark sensors to measure the time-varying electrical field, namely the EFIELD experiment which is part of the DraGMet (Dragonfly Geophysics and Meteorology) package \citep{barnes2021science}. Using two spherical electrodes mounted at different locations on the body of the Dragonfly drone, EFIELD will passively record the AC electrical field between $\sim5$ and \SI{100}{\Hz} with a much finer spectral resolution than PWA/HASI.

In this paper, we describe the numerical model we have developed to simulate Titan’s electromagnetic cavity and predict its resonant frequencies and associated quality factors (section 2). This model is used to build a much less computationally expensive surrogate model which allows to perform an accurate sensitivity analysis of the SR characteristics to the cavity parameters. The surrogate model is then used to re-examine the PWA/HASI measurements leading to results very different from the ones published in Béghin et al. (2012) (section 3). In section 4, it is used to investigate the expected performance of the EFIELD/DraGMet experiment. Lastly, we conclude and discuss the implications of this work in section 5.

\begin{figure}[htb]
\centering

\includegraphics[width=0.7\linewidth]{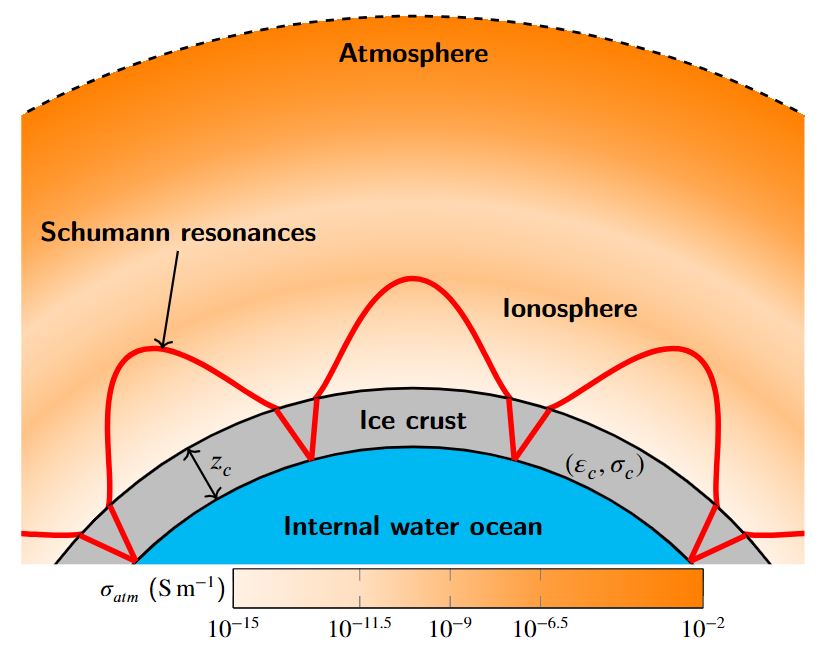}

\caption{Structure and parameters of Titan’s cavity}
\label{fig:SR_Titan}
\end{figure}

\section{Modeling Titan's resonant cavity}
\label{sec:model}

Following \cite{simoes2007new,simoes2008schumann,simoes2008electromagnetic} who developed cavity models for Titan, Venus and other planetary environments, we used the COMSOL Multiphysics© tool to build a numerical model of electromagnetic wave propagation in the cavity of Titan. We then used the numerical model to construct a surrogate model of the propagation of SR on Titan in order to conduct eigenfrequency analysis at a low computing and memory cost.

\subsection{Numerical model}

\subsubsection{Cavity description and parameters}

\begin{figure}
    \centering
    \includegraphics[width=0.6\linewidth]{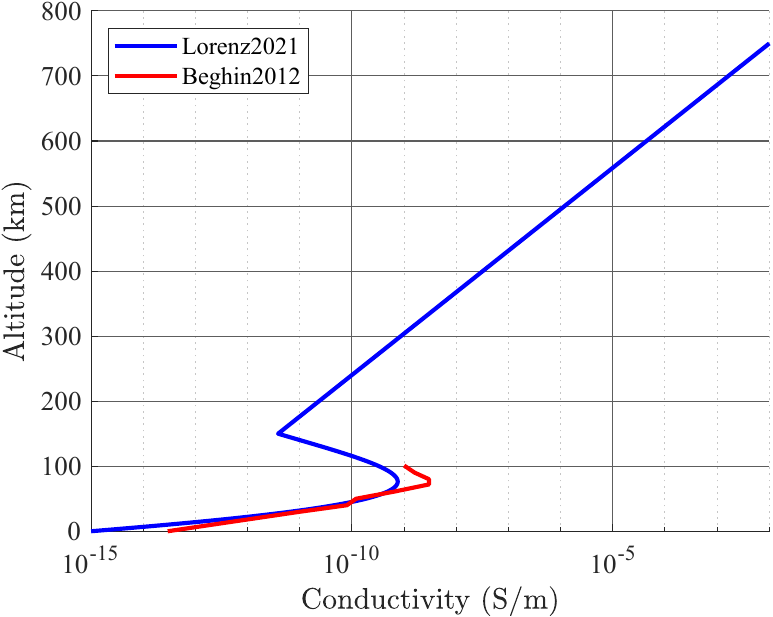}
    \caption{Input conductivity profiles of the atmosphere/ionosphere layer used in the numerical model of Titan’s cavity from \cite{beghin2012analytic}  (red) and \cite{lorenz2021low} (blue)}
    \label{fig:conductivity_profile_beghin}
 \end{figure}

The numerical code solves the Maxwell's equations in a spherical structure made of discrete slabs. Figure \ref{fig:SR_Titan} displays the simplified structure of Titan’s cavity we considered; it consists of three concentric layers:

\begin{itemize}
    \item \textit{the atmosphere/ionosphere layer} for which an analytic conductivity profile is given to the model, namely the one proposed by \cite{beghin2012analytic} or, more recently, by \cite{lorenz2021low}  displayed in figure \ref{fig:conductivity_profile_beghin}. Both conductivity models include an ionized layer at an altitude of about $60-70$  \si{\km} on which ELF waves are reflected. They rely on Huygens measurements of the electron-density performed from an altitude of \SI{140}{\km} down to the surface \citep{grard2006electric}. In \cite{lorenz2021low}, the conductivity profile is interpolated from $140$ to \SI{750}{\km} and better respect the upper limit of the near-surface conductivity imposed by Huygens Relaxation Probe measurements.
    \item \textit{the ice crust layer} in which ELF waves are refracted following Fresnel’s laws. This layer is assumed uniform in terms of electrical properties with a very small conductivity that allows ELF waves to propagates over a very long path (i.e., the skin depth is $>\SI{1000}{\km}$).
    \item \textit{the salty ocean layer} which is assumed to be a perfect electric conductor and therefore on which ELF waves are fully reflected.
\end{itemize}

The parameters of the cavity model considered for parametric analysis are the followings:
\begin{itemize}
    \item \textit{the thickness of the ice crust $z_c$} : Based on Cassini and Huygens observations as well as on gravity and thermal modeling (see section \ref{sec:introduction}), we consider that $z_c$ can vary over a wide range of values from $5$ to \SI{200}{\km}, and most likely between $40$ and \SI{170}{\km}.

    \item \textit{the real part of the ice crust relative permittivity $\varepsilon_c$}: The relative permittivity of water ice at Titan’s temperatures and ELF frequencies is $\sim3$ (e.g., \cite{mattei2014dielectric}). However, the crust permittivity also depends on its porosity and on the presence of impurities or contaminants such as ammonia. We therefore test values in the range $2-4$ which encompasses the value measured at the Huygens landing site by the permittivity probe PWA-MIP/HASI, namely $2.5\pm0.3$ \citep{hamelin2016electrical}.

    \item \textit{the electrical conductivity of the ice crust $\sigma_c$}: Through expected to be small, the conductivity of the ice crust of Titan remains uncertain. It is especially sensitive to the possible presence of ionic contaminants. \cite{beghin2012analytic} investigates the $1-4$ \si{\nano\siemens\per\meter} range while \cite{hamelin2016electrical} found a conductivity of $1.2\pm0.6$ \si{\nano\siemens\per\meter} at the Huygens landing site. To account for this measurement, we consider values in the range $0.6-4$ \si{\nano\siemens\per\meter}.
\end{itemize}

As outputs, the numerical model computes the eigenfrequencies of the first three modes of the resonant cavity $(f_1,f_2,f_3)$ along with their corresponding Q-factors (quality factors: $Q_1,Q_2,Q_3)$ which describe wave attenuation in the cavity. More specifically, the model provides the complex frequencies of the different eigenmodes from which the Q-factor is computed as followed:
\begin{equation}
    Q_n=\frac{\mathrm{Re}(f_n)}{2\mathrm{Im}(f_n)}\approx\frac{f_n}{\Delta f_n}
    \label{eq:q_factor}
\end{equation}
where $\mathrm{Re}$ and $\mathrm{Im}$ are respectively the real and imaginary parts of the complex eigenfrequency, $f_n$ is the peak power frequency of mode $n$, and $\Delta f_n$ is the width at half-power.

For a given uncertainty $\delta f_n$ on $f_n$ and $\Delta f_n$, which in practice is mainly dictated by the measurement spectral resolution, the uncertainty $\delta Q_n$ on the corresponding quality factor $Q_n$ can be derived by logarithmic differentiation of equation \ref{eq:q_factor}:
\begin{equation}
    \delta Q_n \approx Q_n (1+Q_n)\frac{\delta f_n}{f_n}
    \label{eq:uncertainty_q_factor}
\end{equation}

\subsubsection{Numerical approach}

The numerical model uses the Finite Element Method \citep{zimmerman2006multiphysics} for solving Maxwell’s equations with the boundary conditions and layers properties as described above. Since layers properties are only functions of the radial distance, the resonant cavity problem can be solved in a 2D axi-symmetric configuration. We nevertheless validate our 2D model with comparison to a 3D model and results from the numerical model from \cite{simoes2007phd}.

\begin{table}[h]

\caption{Comparison of results with the 2D axi-symmetric approximation model with the 3D model for two different cavities (and associate relative error)} \label{tab:comparison_2D_3D}

\begin{subtable}{0.49\textwidth}
\centering
   \begin{tabular}{ccccc}
      \toprule
       Model & $Q_1$ & $\Delta Q_1/Q_1$ & $f_1$ & $\Delta f_1/f_1$  \\ 
      \midrule
      2D axi-symmetric & 3.20 & 4.69\% & \SI{32.61}{\Hz} & 0.797\%  \\ 
      3D & 3.05 & 4.92\% & \SI{32.35}{\Hz} & 0.803\%  \\ 
      \bottomrule
   \end{tabular}
   \bigskip
   \caption{First case: $z_c=\SI{20}{\kilo\meter},\varepsilon_c=3.94,\sigma_c=\SI{9.69e-9}{\siemens\per\meter}$}
\end{subtable}
\hfill
\begin{subtable}{0.49\textwidth}
\centering
   \begin{tabular}{cccc}
      \toprule
       $Q_1$ & $\Delta Q_1/Q_1$ & $f_1$ & $\Delta f_1/f_1$  \\ 
      \midrule
      2.735 & 2.93\% & \SI{18.28}{\Hz} & 1.70\%  \\ 
      2.655 & 3.01\% & \SI{17.97}{\Hz} & 1.73\%  \\ 
      \bottomrule
   \end{tabular}
      \bigskip
   \caption{Second case: $z_c=\SI{15}{\kilo\meter},\varepsilon_c=2.5,\sigma_c=\SI{1e-9}{\siemens\per\meter}$}
\end{subtable}
\end{table}

\begin{table}[h]
\caption{Comparison of results with the 2D axi-symmetric approximation model with the model used in \cite{simoes2007phd} (relative error $\varepsilon$) for a lossless atmosphere} \label{tab:comparison_2D_simoes}
\centering
   \begin{tabular}{ccccccc}
      \toprule
       Model & $f_1$ & $\Delta f_1/f_1$ & $f_2$ & $\Delta f_2/f_2$ & $f_3$ & $\Delta f_3/f_3$  \\ 
      \midrule
      2D axi-symmetric & \SI{22.54}{\Hz} & 1.06\% & \SI{39.08}{\Hz} & 1.20\% & \SI{55.27}{\Hz} & 1.23\%  \\ 
      \cite{simoes2007phd} & \SI{22.30}{\Hz} & 1.08\% & \SI{38.61}{\Hz} & 1.22\% & \SI{54.59}{\Hz} & 1.25\%  \\ 
      \bottomrule
   \end{tabular}

\end{table}

\begin{figure}[htb]
\centering
\begin{subfigure}{0.49\textwidth}
    \centering
    \includegraphics[width=0.6\linewidth]{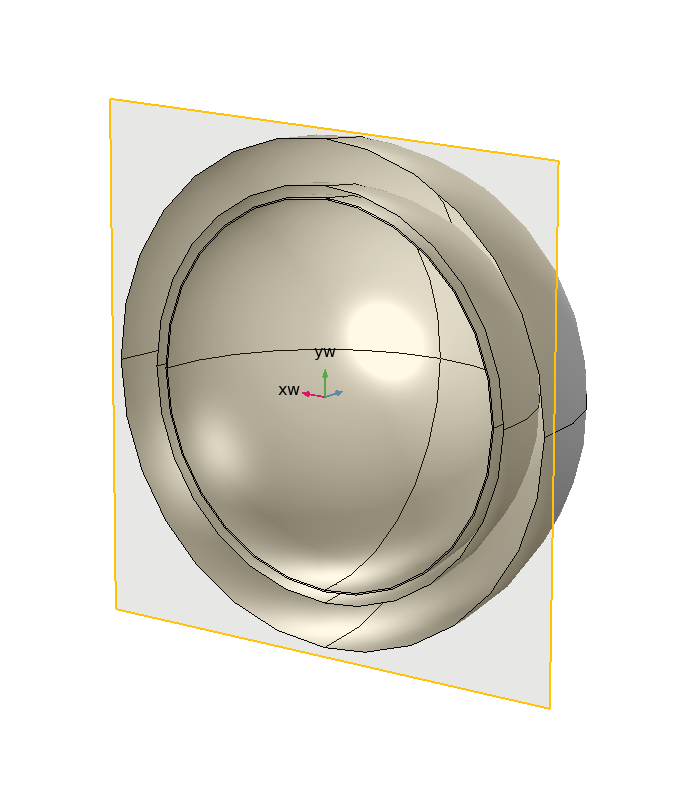}
    \caption{}
\end{subfigure}
\hfill
\begin{subfigure}{0.49\textwidth}
    \centering
    \includegraphics[width=\linewidth]{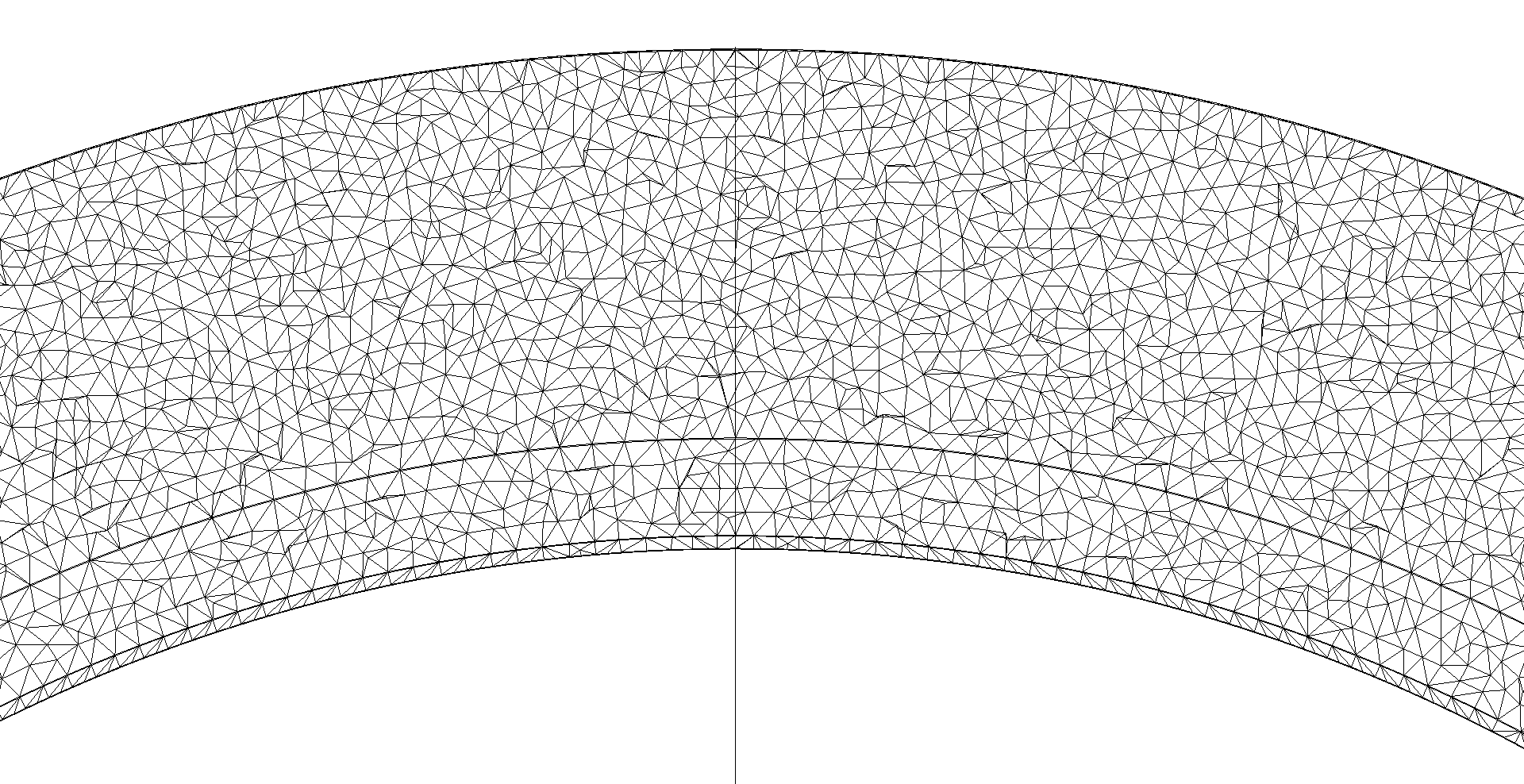}
    \caption{}
\end{subfigure}      
\caption{(a) 3D model of Titan’s planetary cavity on COMSOL Multiphysics, (b) 2D cut of the 3D mesh on (a).}
\label{fig:titan_model_comsol}
\end{figure}

An example of a 3D model of Titan’s cavity is displayed on figure \ref{fig:titan_model_comsol} as well as a 2D cut of the mesh. Due to the level of discretization needed to accurately reproduce the behavior of the electric field in Titan's atmosphere, the complete 3D mesh consists of $774,258$ domain elements, $181,100$ boundary elements, and $2,588$ edge elements. A single resolution of Maxwell's equations using such a mesh takes $\sim\SI{2}{\hour}$ (on an Intel Core i5-12500H, \SI{2.5}{\giga\Hz}, $32$ GB of RAM). In contrast, the design of a 2D axi-symmetric model with a mesh composed of $202,993$ domain elements and $3,652$ boundary elements requires $\sim\SI{30}{\s}$ of computation time which is much more reasonable for the optic of using the model to perform an accurate parametric inversion.

Table \ref{tab:comparison_2D_3D} reports the results from the 3D model and the 2D approximation for two different cavities. The 2D approximation is accurate enough so that the results from the 3D model are reproduced with relative errors smaller than $5\%$ for both the Q-factor and the resonant frequency. Table \ref{tab:comparison_2D_simoes} reports the results from the 2D approximation and the model from \cite{simoes2007phd} for a lossless atmosphere showing that the 2D approximation is able to accurately reproduce the cavity behavior in the case of study case with a relative error on the first three resonant frequencies smaller than $2\%$. Thus, for the remainder of the paper, we only consider the 2D axi-symmetrical model of Titan's planetary cavity.

\subsection{Surrogate model}

Even with the 2D axi-symmetrical approximation, a complete sensitivity analysis or data inversion using Titan's cavity numerical model would require to compute several thousand of datapoints. To avoid this, we built a surrogate model (also called "metamodel") which provides an approximate description of the behavior of Titan’s resonant cavity by analytical functions. The resulting metamodel can then be used, instead of the numerical model, to perform various analysis (e.g. optimization, sensitivity analysis) at a low computational cost \citep{van2016sensitivity}.

\subsubsection{Description}

The surrogate model used here is an exact interpolator; the Polynomial-Chaos based Kriging (PCK) \citep{schobi2015polynomial} metamodeling combines both Polynomial Chaos Expansion (PCE) and Kriging to predict the variations of a given model $\mathcal{M}(X)$. Kriging is used to interpolate the local variations of the output model while PCE is useful for the global approximation. A PCK metamodel is defined by:
\begin{equation}
    \mathcal{M}(X)= \sum\limits_{\boldsymbol\alpha\in\mathcal{A}}y_{\boldsymbol\alpha}\psi_{\boldsymbol\alpha}(X) + \sigma^2 Z(X,\omega)
    \label{eq:pck}
\end{equation}
where $\sum_{\boldsymbol\alpha\in\mathcal{A}}y_{\boldsymbol\alpha}\psi_{\boldsymbol\alpha}(X)$ is a weighted sum of orthonormal polynomials describing the trend of the PCK model and $\sigma^2 Z(X,\omega)$ is a zero-mean stationary Gaussian process with a variance of $\sigma^2$. The computation of the metamodel parameters is performed by the UQLab framework available on Matlab \citep{marelli2014uqlab}.

The costly part of the metamodelling process is the training time which can be greatly reduced by using sequential sampling instead of classical space-filling approaches. In this work an adaptive sampling algorithm combined with PCK has been used to build the metamodel. This algorithm has already been proven useful for various electromagnetic problems \citep{lagouanelle2023adaptive}.

\subsubsection{Metamodel accuracy}
\label{se:metamodel_accuracy}

Once built, the surrogate model is used to predict the behavior of the cavity outside of the training data. For such a purpose, a proper metric is crucial to quantify the accuracy of these predictions. A classical approach consists in using a validation dataset outside of the training dataset and computing the mean squared error (MSE) of the metamodel prediction compared to the real input values. However, the resulting MSE is biased by the use of only one dataset and could vary greatly from one validation dataset to another. Moreover, this approach requires additional calls of the expensive computational numerical model for building the training dataset, which ultimately increases the computation time.
Thus, a better metric was chosen : the Leave-one-out cross-validation error ($LOO$), which does not require additional computations. 

Let us consider a set $\{(X_1,Y_1),\ldots,(X_N,Y_N)\}$ of $N$ input samples. Using this set, one can build a PCK metamodel $\mathcal{M}$ and evaluate the $LOO$ as follows:
\begin{equation}
     LOO= \frac{1}{N}\sum_{i=1}^{N} \left ( \dfrac{||\mathcal{M}_{/i}(X_i)-Y_i||}{||Y_i||} \right ) ^2
     \label{eq:loo}
\end{equation}
where $\mathcal{M}_{/i}$ is the mean predictor that was trained using all $(X, Y)$ except $(X_i, Y_i)$.
For a given datapoint $(X_i,Y_i)$, a metamodel is built with all datapoints except datapoint $i$, which gives $N-1$ training datapoints.  This metamodel is then used to predict the value $Y_i$ at the remaining datapoint $i$, where the difference is classified with a MSE. The process is repeated for every datapoint which, after average, provides the $LOO$. 
The use of $N$ different validation sets of one datapoint guarantees that the $LOO$ is much less biased than a classical MSE and reduces the probability of overestimating the validation error \citep{elisseeff2003leave}. In this work, we therefore consider the $LOO$  as our accuracy metric. A $LOO$  close to $1$ (or $100\%$) implies that the surrogate model does not provide a good approximation of the system. On the other hand, the smaller the $LOO$, the more accurate the surrogate model.

\subsubsection{Sensitivity indices}
\label{sec:sobol_indices}

Since the resulting metamodel consists of an analytical function, calling the metamodel is extremely cheap in terms of computation time. Thus, sensitivity analyses, which are usually performed by Monte-Carlo analyses over the parameter spaces, are now feasible at a low computation cost. 

The sensitivity analysis we conduct relies on Sobol' indices which are scalars between $0$ and $1$ describing the influence of a set of inputs on a model output \citep{sobol1993sensitivity}. The most commonly used Sobol' indices is the first-order Sobol' index defined, for a given parameter $P_i$, as:
\begin{equation}
    S_i=\frac{\mathrm{Var}_{P_i}(\mathbb{E}_{\boldsymbol X_{/i}}(\boldsymbol Y | X_i))}{\mathrm{Var}(\boldsymbol Y)}
    \label{eq:first_order_sobol_index}
\end{equation}
$S_i$ is a measure of the fraction of the output variance caused by the variance of a given input parameter. In other words, it describes the impact of a parameter $P_i$ alone on the output model compared to other parameters. The closer to 1, the bigger impact $P_i$ has on the model output.

However,  parameters are usually not independent and their relative effects cannot be separated from each other. This leads to the definition of higher-order Sobol' indices as, for a subset of parameters $(P_{i_1},...,P_{i_s})$:
\begin{equation}
    S_{i_1,...,i_s}=\frac{\mathrm{Var}_{(P_{i_1},...,P_{i_s})}(\mathbb{E}_{\boldsymbol X_{/i_1...i_s}}(\boldsymbol Y | X_{i_1},...,X_{i_s}))}{\mathrm{Var}(\boldsymbol Y)}
    \label{eq:higher_order_sobol_index}
\end{equation}
which describes the sensitivity of the model to the variations of several input parameters simultaneously.

For high dimensional output models, the interpretation of all orders Sobol' indices can be difficult due to the high number of possible combinations. Therefore, for an input parameter $P_i$, a total-effect index (or total Sobol' index) $S_i^T$ is defined by summing all the Sobol' indices as follows:
\begin{equation}
    S_i^T= \sum\limits_{\{\boldsymbol u, \boldsymbol u\subseteq \llbracket 1,d \rrbracket \text{ and } i\in \boldsymbol u \}} S_{\boldsymbol u}
    \label{eq:total_sobol_index}
\end{equation}
$S_i^T$ is the most suited sensitivity tracker for our study and will be refereed to as $S_i$ in the remainder of the paper. When using PCK metamodels, the computation of the various Sobol' indices of the surrogate model can be easily extracted from the polynomial decomposition (see equation \ref{eq:pck}). Therefore, no additional computation of the surrogate model are required to perform the sensitivity analysis based on the total Sobol' indices.

\section{Re-assessment of PWA/HASI/Huygens observations}
\label{sec:huygens_analysis}

\cite{fulchignoni2005situ} first reported the detection of a narrow power line at $\sim\SI{36}{\Hz}$ in the ELF spectrum measured by PWA/HASI during Huygens’ descent in Titan’s atmosphere in January $2005$, from an altitude of \SI{140}{\km} down to the surface. The magnitude of this signal is especially enhanced just after the deployment of the stabilizer parachute, at an altitude of $\sim\SI{110}{\km}$. \cite{beghin2007schumann} proposed different scenarios, both natural and artificial, to explain the \SI{36}{\Hz} signal. In \cite{beghin2012analytic}, a natural scenario is preferred: the signal and associated Q-factor of about $\sim6$ would be the second harmonic of a SR propagating between Titan’s ionosphere and ocean and triggered by interactions with Saturn’s magnetosphere. Using an approximate analytical model of Titan’s cavity, \cite{beghin2012analytic} further derive constraints on the physical parameters of the cavity from PWA/HASI measurements. More specifically, they conclude that the measured $f_2=36\pm\SI{3}{\Hz}$ and $Q_2\sim6$ are indicative of a water-ammonia ocean lying at a depth of $40-\SI{80}{\km}$.

In this section, we re-asses the PWA/HASI data using the surrogate model we have developed (see section \ref{sec:model}) to investigate, in a more accurate fashion, the constraints Huygens measurements bring on the thickness of the ice crust (i.e., the depth of the ocean) $z_c$. As a starting point, we adopt exactly the same hypotheses as in \cite{beghin2012analytic} that is the same conductivity profile in the atmosphere and ranges of variation for $\varepsilon_c$ ($2-4$) and $\sigma_c$ ($1-4$ \si{\nano\siemens\per\meter}).

The metamodelling process estimates consistently ($LOO\approx3.9\%$, $n_{samples}=1584$) the second harmonic of SR ($f_2$, $Q_2)$. Using the resulting surrogate model, two regular 3D grids of $f_2$ and $Q_2$ values along all three input parameters $z_c$, $\varepsilon_c$ and $\sigma_c$ can be computed at a low computation cost: $50\times50\times50=125,000$ values, which would have taken $43$ days of computation time using directly the numerical model instead of only $13$ hours for training the metamodel. 

\begin{figure}[htb]
    \centering
    
 \end{figure}

\begin{figure}[htb]
     \centering
     \begin{subfigure}[b]{0.49\linewidth}
     \centering
   \includegraphics[width=\linewidth]{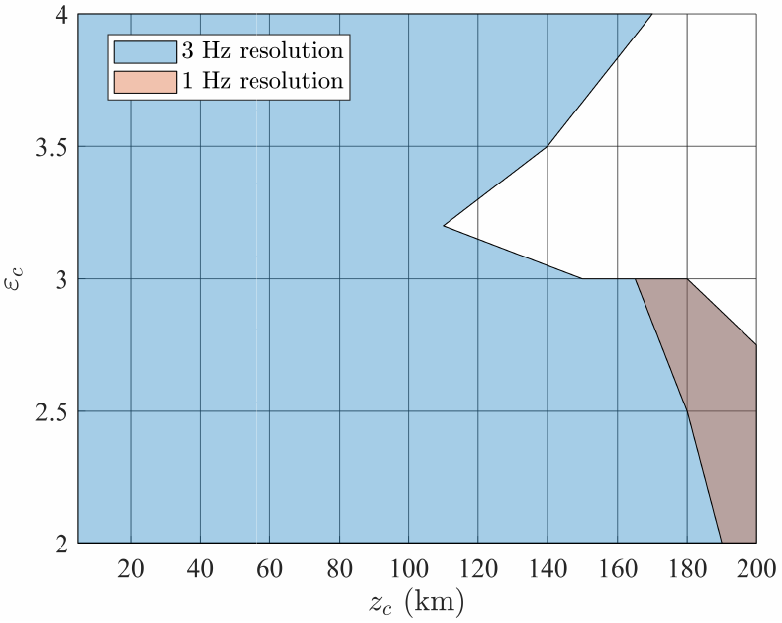}
    \caption{}
    \label{fig:inversion_huygens_data_beghin}
    \end{subfigure}
     \hfill
     \begin{subfigure}[b]{0.49\linewidth}
         \centering
    \includegraphics[width=\linewidth]{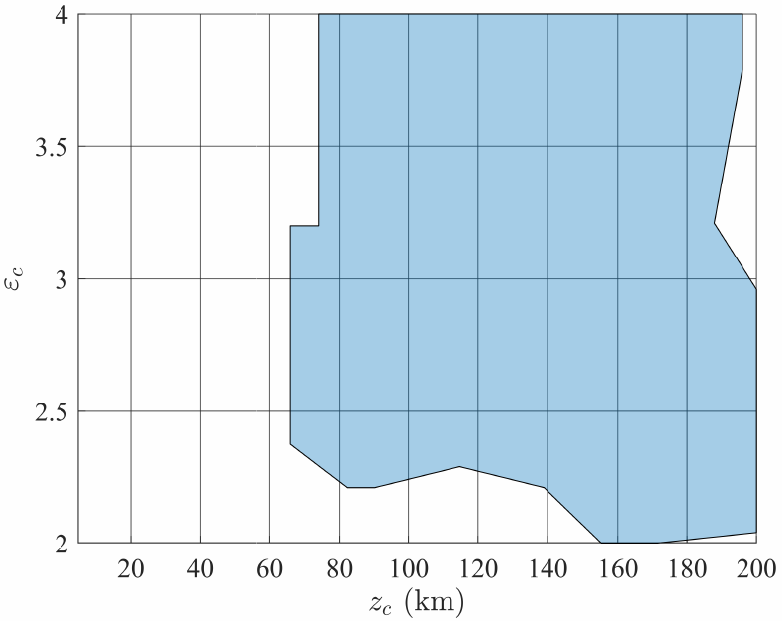}
    \caption{}
    \label{fig:inversion_huygens_data_beghin_first}
     \end{subfigure}
        \caption{Inversion of Huygens PWA/HASI data taking the same hypotheses as in \cite{beghin2012analytic} (i.e., conductivity profile, $\varepsilon_c\in[2-4]$ and $\sigma_c\in[1-4]$ \si{\nano\siemens\per\meter}) and using the surrogate model developed in this work (a). All values of $z_c$ can be in a combination with $\varepsilon_c$ and $\sigma_c$ that reproduces Huygens data which were measured with a \SI{3}{\Hz} spectral resolution. A \SI{1}{\Hz} spectral resolution would drastically reduce the parameter space for $z_c$. (b) displays the same inversion exercise but considering that the detected line at \SI{36}{\Hz} is the fundamental of the SR instead of the second harmonic.}
\end{figure}

By analyzing the 3D grids, every combination of $z_c$, $\varepsilon_c$ and $\sigma_c$ which gives $f_2\in[33-39]\si{\Hz}$ and $Q_2\in[3-9]$ can be classified as a potential solution explaining Huygens measurements. Figure \ref{fig:inversion_huygens_data_beghin} displays all the potential inversion of Huygens measurements in the plane $(z_c,\varepsilon_c)$ where all the grids in the direction $\sigma_c$ have been stacked. When accounting for \SI{3}{\Hz} spectral resolution, the range of possible values for $z_c$ covers all the parameter space (i.e. $5-\SI{200}{\km}$) which means that no constraint can be deduced on $z_c$ from the PWA/HASI dataset (blue zone). On the other hand, if the signal characteristics were known with a \SI{1}{\Hz} resolution, only a narrow range of $z_c$ values could explain the observations (pink zone). This hypothetical inversion would restrain the thickness of the ice crust to $z_c\in [165-200]$ \si{\km}.

Our results are in contraction with \cite{beghin2012analytic} which can be explained by the various analytical approximations they consider to solve the wave propagation equation. Furthermore, since the metamodel is sufficiently accurate ($LOO\approx3.9\%$), it can be used to conduct an accurate Sobol'-based sensitivity analysis. The following total Sobol' indices are found: $S_{z_c}=0.73$, $S_{\varepsilon_c}=0.28$ and $S_{\sigma_c}=0.74$. These indexes indicate that the ice crust thickness is the parameter that has the most significant impact on the SR characteristics. However, since all indexes are of the same order of magnitude, no parameter can be regarded as having a negligible impact. This is a further guarantee that the metamodel accuracy is correctly estimated with the $LOO$ and that our metamodel is highly accurate. Therefore the differences from    \cite{beghin2012analytic} 's results and the present work  cannot be ascribed to an incorrect estimate of the errors from the metamodel.

The surrogate model also demonstrates that solutions (in terms of combinations of $z_c$, $\varepsilon_c$ and $\sigma_c$ values) in which the detected signal is not the second harmonic but the fundamental are possible (figure \ref{fig:inversion_huygens_data_beghin_first}). If so, the constraints on $z_c$ derived from PWA/HASI would be different (namely >\SI{80}{\km}, see figure \ref{fig:inversion_huygens_data_beghin_first}).

\begin{figure}[htb]
    \centering
    \includegraphics[width=0.6\linewidth]{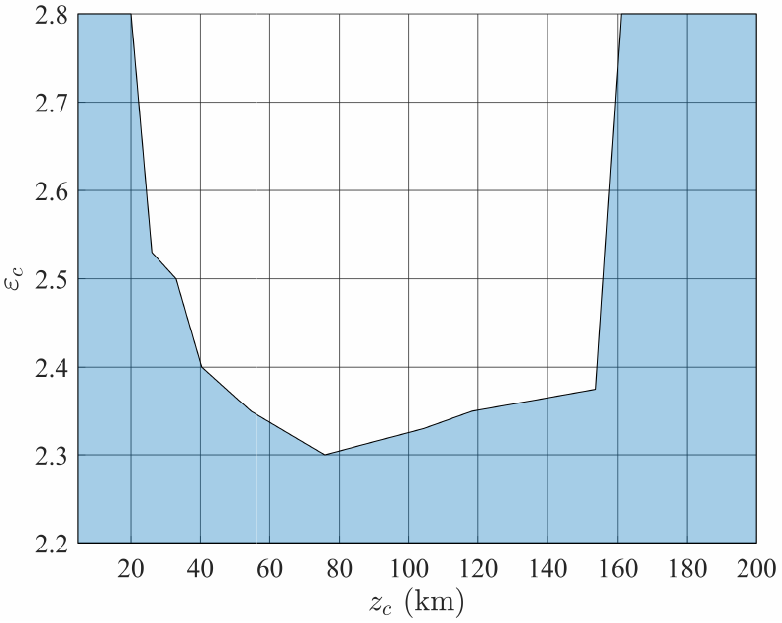}
    \caption{Inversion of Huygens PWA/HASI  using the surrogate model developed in this work with up-to-date assumptions, namely the atmosphere conductivity profile from \cite{lorenz2021low} and $\varepsilon_c\in[2.2-2.8]$ and $\sigma_c\in[0.6-1.8]$ \si{\nano\siemens\per\meter} as estimated by \cite{hamelin2016electrical}.}
    \label{fig:inversion_huygens_data_ralph}
\end{figure}

Lastly, figure \ref{fig:inversion_huygens_data_ralph} shows how the constraints on $z_c$ are modified if the numerical model from which the surrogate model was built rather uses \cite{lorenz2021low}'s conductivity profile and values of $\varepsilon_c$ and $\sigma_c$ from \cite{hamelin2016electrical}. We will keep these hypotheses for the remainder of the paper and, in particular, to assess EFIELD/DraGMet future performance in the frame of the Dragonfly mission. The associated sensitivity analysis provides the following total Sobol' indices: $S_{z_c}=0.78$, $S_{\varepsilon_c}=0.34$ and $S_{\sigma_c}=0.51$. Again and notably, $z_c$ is the  parameter that have the most impact on SR characteristics. Nevertheless, the effects of $\varepsilon_c$ and $\sigma_c$ cannot be neglected as their respective total Sobol' index are of the same order of magnitude. This further implies that their accurate knowledge would greatly reduce the uncertainty on the inversion of Huygens measurements as well as be very valuable for the analysis of future EFIELD data. In particular, if the real part of the permittivity of the ice crust were measured as $\varepsilon_c=2.5\pm0.1$ instead of $2.5\pm0.3$, it would help discriminating between the two distinctive domains in figure \ref{fig:inversion_huygens_data_ralph}: $z_c\in[5-40]\si{\km}$ and $z_c\in[150-200] \si{\km}$. This is further discussed below.

\section{Anticipated EFIELD/DraGMet/Dragonfly performance}

\label{sec:efield_experiment}

This section investigates the performance and possible outcome from the forthcoming electric-field experiment on Titan.

\subsection{The EFIELD experiment on board Dragonfly} 

In June 2019, NASA selected the Dragonfly mission project for its New Frontiers program \citep{turtle2018dragonfly,lorenz2018dragonfly,barnes2021science}. The primary goal of the Dragonfly mission is to investigate the chemistry and habitability of Titan. Starting operations in mid-2030s, the Dragonfly quadcopter drone will visit a variety of sites, from a dune field to the rim of a young impact crater, and sample materials in different geologic settings. Dragonfly embarks a Geophysical and Meteorological package (DraGMet) which is a suite of sensors designed to measure e.g. the temperature, pressure, methane humidity, wind speed and direction, ground dielectric constant, thermal properties and level of seismic activity at each Dragonfly landing site. 

Among these sensors is the EFIELD experiment which consists of two independent spherical electrodes ($\sim\SI{5}{\cm}$ in diameter) accommodated at The End of $\sim\SI{25}{\cm}$ long stalks pointing away from the drone body. From two different locations on the drone, these electrodes will passively record the time-varying electrical field at low frequencies ($\sim5-100$ \si{\Hz}) with the main goal of detecting SR, if any. As a secondary objective, the EFIELD probes will detect and characterize near-surface wind-blown charged grains flying in their vicinity \citep{chatain2023detection}. 

The EFIELD experiment offer many advantages over the PWA/HASI one. It will operate during an extended period of time (several times a Titan day for the $3.3$ years of the nominal mission), from a stable and much mechanically-quieter platform than the Huygens probe. Further, the EFIELD design should guarantee a spectral resolution of \SI{1}{\Hz} (against, at best, \SI{3}{\Hz} for PWA/HASI) and the capture of the first three harmonics of the SR. Figure \ref{fig:inversion_huygens_data_beghin} demonstrates how valuable a finer spectral resolution would be to bring more robust constraints on the depth of the buried ocean $z_c$. The benefit of detecting more than one SR harmonics is investigated below.

\subsection{Multi-modal analysis}

The re-assessment of Huygens data presented in section \ref{sec:huygens_analysis} relies on the measurement of only one mode of the SR; it concludes that a wide range of values are possible for the thickness of the ice crust ($z_c\in[5-200]$ \si{\km}). The multi-modal analysis of SR enabled by EFIELD should drastically reduce this range because, in addition to a better spectral resolution, the three first modes of the Schumann resonances will be associated with three different domains of possible $z_c$ values wh intersection may be narrow.

\begin{figure}[htb]
     \centering
     \begin{subfigure}[b]{0.49\linewidth}
     \centering
    \includegraphics[width=\linewidth]{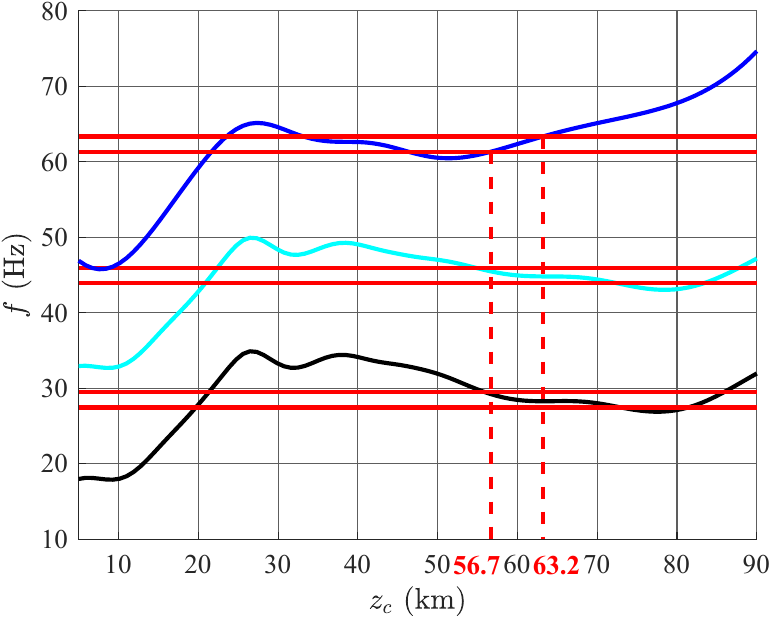}
    \caption{}
    \label{fig:efield_expected_performances_multimodal_inversion_frequency}
    \end{subfigure}
     \hfill
     \begin{subfigure}[b]{0.49\linewidth}
         \centering
        \includegraphics[width=0.97\linewidth]{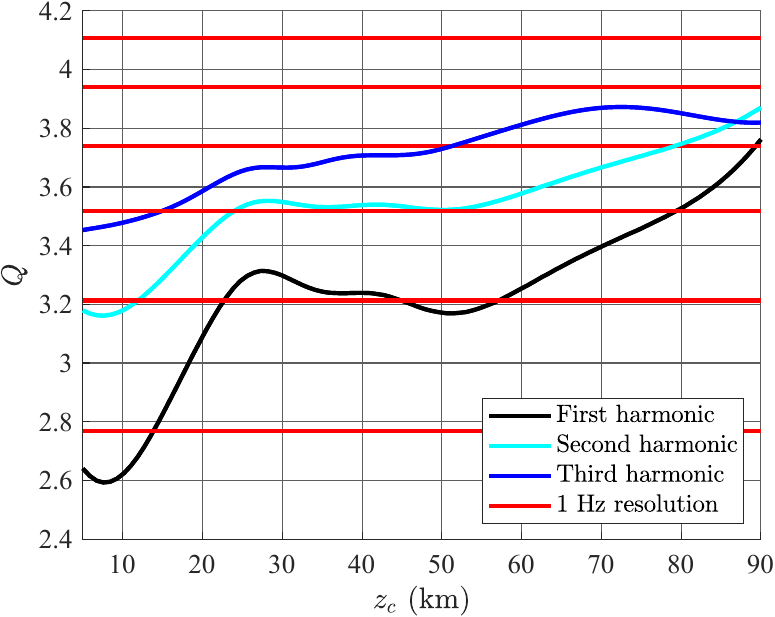}
         \caption{}
        \label{fig:efield_expected_performances_multimodal_inversion_q_factor}
     \end{subfigure}
        \caption{Variations of the resonant frequencies ($f_1,f_2,f_3$) (a) and of the quality factors ($Q_1,Q_2,Q_3$) (b) as a function of the thickness of the ice crust $z_c$ assuming $\varepsilon_c=2.5,\sigma_c=\SI{1.2}{\nano\siemens\per\meter}$. In a case where the input value is $z_c=\SI{60}{\km}$, the multi-modal analysis provides a range of values for $z_c$ that is between \SI{56.7}{\km} and \SI{63.2}{\km} (dashed red lines)}
        \label{fig:efield_expected_performances_multimodal_inversion}
\end{figure}

As an illustration, we numerically simulate Titan’s cavity for the following parameter values: $z_c=\SI{60}{\kilo\meter},\varepsilon_c=2.5,\sigma_c=\SI{1.2}{\nano\siemens\per\meter}$ and the up-to-date conductivity profile from \cite{lorenz2021low}. The numerical model provides the following outputs: $f_1=28.4$  \si{\Hz},  $f_2=44.9$ \si{\Hz},  $f_3=62.4$ \si{\Hz} along with their corresponding quality factors: $Q_1=3.25$, $Q_2=3.58$ and $Q_3=3.81$. Using the metamodel, we further compute the variations of the resonant frequencies ($f_1,f_2,f_3$) and of the quality factors ($Q_1,Q_2,Q_3$) as a function of the thickness of the ice crust $z_c$, assuming fixed values for $\varepsilon_c=2.5$ and $\sigma_c=\SI{1.2}{\nano\siemens\per\meter}$. These variations are displayed on figure \ref{fig:efield_expected_performances_multimodal_inversion}.

\label{sec:multi-modal analysis}
 
Assuming a \SI{1}{\Hz} resolution, each harmonic can be inverted separately (using the method described in section 3), resulting in three different possible domains for $z_c$:
\begin{itemize}
    \item $f_1=28.4\pm\SI{1}{\Hz}\implies z_c\in[19.7-21.3]\cup[56.0-73.0]\cup[81.4-85.8]\si{\kilo\meter}$
    \item $f_2=44.9\pm\SI{1}{\Hz}\implies z_c\in[20.9-22.4]\cup[54.8-72.3]\cup[83.4-87.8]\si{\kilo\meter}$
    \item $f_3=62.4\pm\SI{1}{\Hz}\implies z_c\in[21.7-23.5]\cup[33.2-46.3]\cup[56.7-63.2]\si{\kilo\meter}$
\end{itemize}
The intersection of these domains is: $z_c\in[56.7-63.2]\si{\kilo\meter}$ which corresponds to an uncertainty of $6\%$ with respect to the input value of $60$ km. A similar inversion is performed on the quality factors separately:
\begin{itemize}
    \item $Q_1=3.25\pm 0.49 \implies z_c\in[13.8-89.4]\si{\kilo\meter}$
    \item $Q_2=3.58\pm 0.36\implies z_c>\SI{11.9}{\kilo\meter}$
    \item $Q_3=3.81\pm 0.29\implies z_c>\SI{14.9}{\kilo\meter}$
\end{itemize}
In the case considered here, no restriction can be further obtained on $z_c$ from the Q-factor values (see figure \ref{fig:efield_expected_performances_multimodal_inversion_q_factor}). Nevertheless, this example well illustrates the values of measuring several modes of the SR.

\subsection{Inversion uncertainty}

The example above assumes fixed values for both $\varepsilon_c$ and $\sigma_c$. To take into account our imperfect knowledge of the electrical properties of Titan’s ice crust, the uncertainties on these parameters have to be propagated through the inversion to estimate their effects on the derivation of $z_c$.  As another illustration, we numerically simulate Titan’s cavity with a thickness of the ice crust fixed at $z_c=\SI{60}{\km}$. A multi-modal analysis is then performed for every point $(z_c,\varepsilon_c,\sigma_c)$ in a regular parameter grid with $\varepsilon_c$ and $\sigma_c$ respectively varying in $2.5\pm0.3$ and $1.2\pm0.6$ \si{\nano\siemens\per\meter} as found at the Huygens landing site by \cite{hamelin2016electrical}. The parameter grid consists of $17\times25\times25=15,300$ datapoints in the 3D parameter space $\{z_c\}\times\{\varepsilon_c\}\times\{\sigma_c\}$. Every datapoint on the grid which returns the desired values for the first three modes considering their measurement uncertainty of \SI{1}{\Hz} is a possible inversion of EFIELD. 

\begin{figure}[htb]
    \centering
    \includegraphics[width=0.6\linewidth]{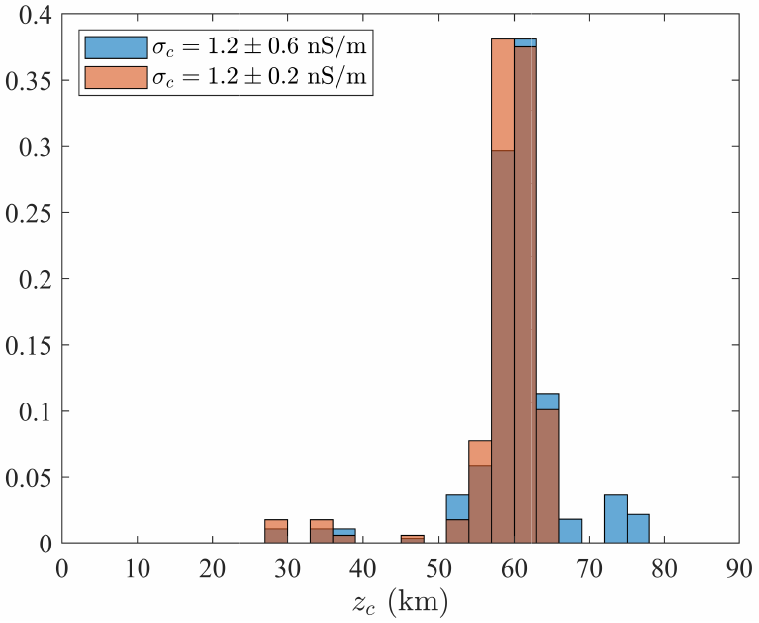}
    \caption{Probability density function of the EFIELD inversion for an ice crust $z_c=\SI{60}{\kilo\meter}$ using the metamodelling process for two different uncertainties on $\sigma_c$}
    \label{fig:expected_performances_efield_example}
 \end{figure}

This allows us to compute the probability density function of the returned values of $z_c$ assuming that the three first harmonic of the SR are detected with a resolution of \SI{1}{\Hz}, as figure \ref{fig:expected_performances_efield_example} displays. Interestingly, $90\%$ of the inversion cases fall in the range $z_c\in[51-69]$ \si{\km} while $\approx10\%$ of the cases return $z_c$ values in the ranges $[20-40]$ \si{\km} or $[70-80]$ \si{\km} (blue bars). Assuming a normal distribution of $z_c$ centered in \SI{60}{\km}, this corresponds to a standard deviation (STD) of \SI{6.8}{\km}. Figure \ref{fig:expected_performances_efield_example} further shows that reducing the uncertainty on $\sigma_c$ from $1.2\pm0.6\si{\nano\siemens\per\meter}$ to $1.2\pm0.2\si{\nano\siemens\per\meter}$ reduces the STD to the value of \SI{5.7}{\km} (orange bars).

\begin{figure}[htb]
     \centering
     \begin{subfigure}[b]{0.49\linewidth}
         \centering
         \includegraphics[width=0.98\linewidth]{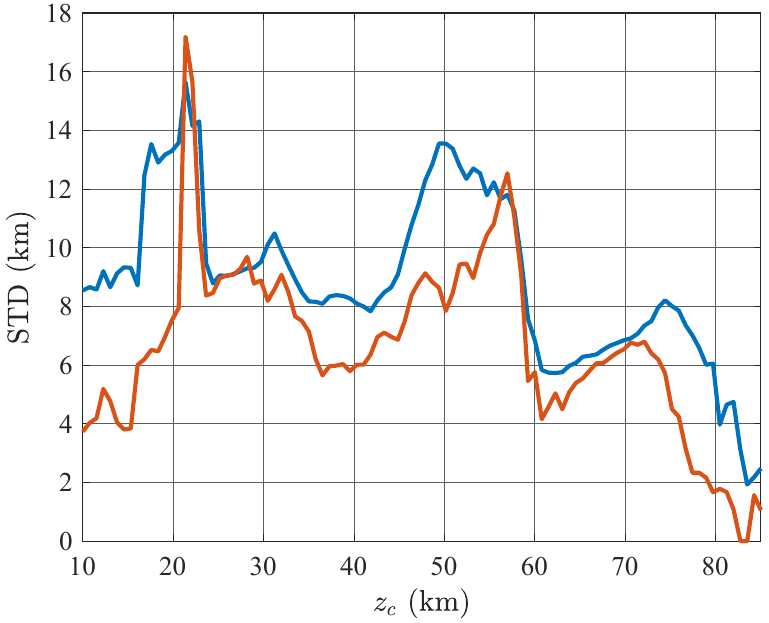}
         \caption{}
         \label{fig:expected_performances_efield_global_uncertainty_km}
     \end{subfigure}
     \hfill
     \begin{subfigure}[b]{0.49\linewidth}
         \centering
         \includegraphics[width=\linewidth]{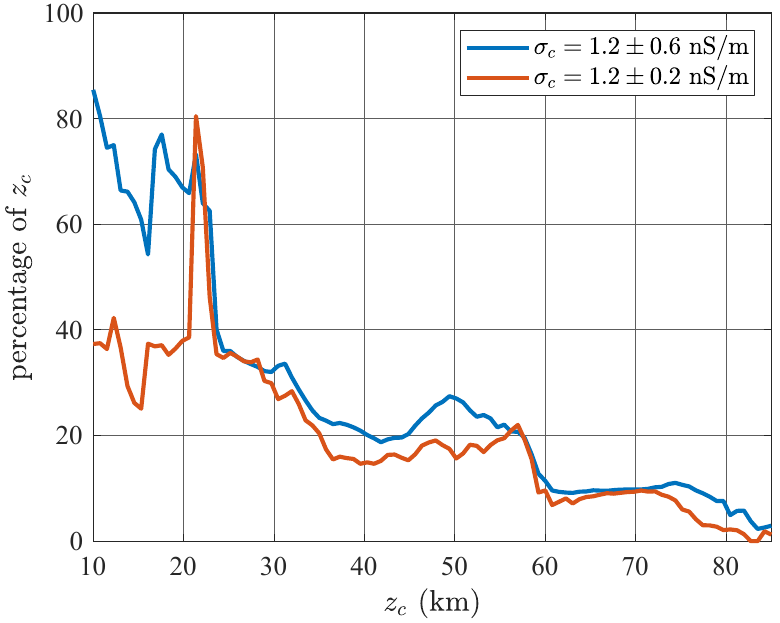}
         \caption{}
         \label{fig:expected_performances_efield_global_uncertainty_percentage}
     \end{subfigure}
        \caption{Standard deviation of the EFIELD inversion against the thickness of the ice crust $z_c$ for two different uncertainties on $\sigma_c$ in \si{\kilo\meter} (a) and percentage of $z_c$ (b)}
        \label{fig:expected_performances_efield_global_uncertainty}
\end{figure}

Such exercise was repeated for a set of $z_c$ values in the $[10-85]$ \si{\km} range to produce figure \ref{fig:expected_performances_efield_global_uncertainty} which displays the STD (in km and $\%$) of the distribution of the inferred $z_c$ as a function of the $z_c$ value for two assumptions on the range of variation of $\sigma_c$. In most cases, the STD is smaller than \SI{10}{\km}. Moreover, it significantly decreases as $z_c$ increases, especially after \SI{60}{\km} where it becomes smaller than $10\%$. In contrast, for small values of $z_c$ ($<\SI{25}{\km}$), the theoretical relative error on $z_c$ can reach almost $100\%$. This can be partially explained by the relative coarse mesh of the numerical model; indeed the dimensions of the cells have been imposed greater than \SI{5}{\km} in order to reduce computation time. Further developments are required to build a surrogate model more appropriate to small thicknesses (i.e., with a finer mesh) but this is out of the scope of this paper.  

The same analysis was conducted assuming a more constrained knowledge of $\sigma_c$ (namely $\sigma_c=1.2\pm0.2\si{\nano\siemens\per\meter}$) resulting in smaller STD for almost all cases (red lines in figure \ref{fig:expected_performances_efield_global_uncertainty}). In particular, the uncertainty at estimating small thicknesses drops from $\approx80\%$ to $\approx40\%$. This further shows the need for a more accurate knowledge of the electrical properties of the crust. This point and others are discussed in the following section.

\section{Discussion and conclusion}

For this work we have developed a numerical model of Titan’s cavity to then build a less computationally expensive surrogate model able to describe how the cavity characteristics (i.e., eigenfrequency and Q-factors) vary with the main cavity parameters (i.e., Titan’s ice crust thickness and electrical properties). This model (and its use for data inversion) is a powerful tool for the analysis of electric field measurements on Titan (and elsewhere). It was used to re-assess Huygens observations leading to the conclusion that the 2005 detection of a line at $\sim\SI{36}{\Hz}$, if indeed due to SR, does not provide any specific constraint on the depth of Titan’s ocean in the range $5-200$ \si{\km} contrary to what is advanced in \cite{beghin2012analytic}. 

The surrogate model was also used to estimate the possible outcomes from the EFIELD/DraGMet/Dragonfly experiment. EFIELD is designed to detect several modes of SR with a fine spectral resolution; we have demonstrated that it has the ability to put a meaningful constraint on the thickness of the ice crust. Considering the electromagnetic properties varying in the ranges specified by \cite{hamelin2016electrical}, the various sensitivity analysis presented throughout this work, reach the same conclusion: although the thickness of the ice crust is the most influential parameter on SR characteristics, the resonant frequencies are dependent in the same order of magnitude on the electrical properties of Titan’s crust. Therefore, in order to reduce the uncertainty at estimating Titan's crust thickness, it is crucial to reduce the uncertainty on the crust electrical properties. 

Another experiment on board the Dragonfly drone will contribute to this task: the DIEL/DraGMet experiment. Acting as a mutual-impedance probe with a pair of electrodes mounted on each skid of the drone, DIEL will measure the complex permittivity (which includes the electrical conductivity) of the ground at several low frequencies ($<\SI{10}{\kilo\Hz}$) thus providing insights into the composition, moisture and porosity of the near-subsurface of Titan as well as on the spatial and temporal variations of such properties. Though all measured permittivity values may not be representative of the crust, values measured on the ejecta blanket of the geologically-young Selk crater (the final destination of Dragonfly) may be. In addition, variations of the measured complex permittivity along the Dragonfly journey to Selk and its possible correlation with otherwise inferred vicinity of the water ice bedrock in the near subsurface will provide further constraint on the ice crust electrical properties. 

Nevertheless, as highlighted in \cite{lorenz2020schumann}, one of the main sources of uncertainty is and will remain our limited knowledge of the lower ionosphere conductivity structure. Unfortunately, no improvement is to be expected from forthcoming observations as Dragonfly will not perform measurements during its descent in Titan’s atmosphere. As a consequence, only theoretical developments can provide further insights on the atmosphere conductivity profile and its expected variations with the local hour, solar activity and the position of Titan in Saturn’s magnetosphere. 

Lastly, future investigations will include the simulation of the actual EFIELD electrodes accommodated on the Dragonfly (conductive) body as well as the study of the effect on measurements of the location and polarization of the possible sources of SR. In the mid-2030s, when the Dragonfly drone will be on Titan, methane-storms are expected at the South Pole which will be the first hypothesis to be tested. Given the different orientation and altitude on the drone of the two EFIELD electrodes, at least two components of the electrical field will be measured. The third component could be captured by rotating the drone and future study will also explore the value of measuring the full electrical field vector.

\section*{Acknowledgements}

The authors wish to thank Ralph Lorenz (APL, PI of DraGMet/Dragonfly) and Jean-Jacques Berthelier, Audrey Chatain and  Franck Montmessin from LATMOS for useful discussions on this project. The authors are also grateful to R\'egion Ile-de-France (DIM-ACAV+ program) for funding this research.

\bibliographystyle{cas-model2-names}




\end{document}